\documentclass{aa}

\usepackage{graphicx}     
\usepackage{siunitx}      
\usepackage{color}        
\usepackage{txfonts}      
\usepackage{longtable}    
\usepackage{mathrsfs}     
\usepackage{lipsum}
\usepackage{subcaption}         
\usepackage{lscape}             
\usepackage{placeins}           
                                
\usepackage[colorlinks=true, linkcolor=blue, citecolor=blue, urlcolor=blue]{hyperref}

\begin{document}

   \title{The dependence of the asteroid rotation on their composition.}

   \subtitle{A spectral class database for MP3C}


 \author{T.~J.~Dyer\inst{1,2}
          \and
          W.-H. Zhou\inst{3}
          \and
          C. Avdellidou \inst{2}
          \and
          M. Delbo \inst{1,2}
          \and
          D. Athanasopoulos \inst{4}
          \and
          J. \v{D}urech \inst{5}
          \and
          P. Pravec \inst{6}
        }

   \institute{Université C\^ote d’Azur, CNRS–Lagrange, Observatoire de la C\^ote d’Azur, CS 34229 – F 06304 NICE Cedex 4, France\\
              \email{t.dyer@oca.eu}
        \and
             University of Leicester, School of Physics and Astronomy, University Road, LE1 7RH, Leicester, UK
        \and 
        JSPS International Research Fellow, Department of Earth and Planetary Science, The University of Tokyo, Tokyo, Japan
        \and
              Institute for Astronomy, Astrophysics, Space Applications and Remote Sensing, National Observatory of Athens, Metaxa \& Vas. Pavlou St., 15236 Penteli, Athens, Greece
        \and
             Charles University, Faculty of Mathematics and Physics, Institute of Astronomy, V Holešovičkách 2, 180 00, Prague, Czech Republic
        \and
            Astronomical Institute, Academy of Sciences of the Czech Republic, Fri\v{c}ova 298, Ond\v{r}ejov, CZ-25165, Czech Republic
             \\
             }

   \date{Received September 30, 20XX}

 
\abstract
{The rotational properties of asteroids provide critical information about not only their internal structure but also their collisional and thermal histories. Previous work has revealed a bimodal distribution of asteroid spin rates, dividing populations into fast and slow rotators, but to date, this separation remains poorly understood (e.g. its dependency on composition).}
{We investigate whether the valley separating fast and slow rotators in rotational period-diameter space depends on the composition of the asteroid. The latter is approximated by asteroids' spectral class.}
{First, we extended the Minor Planet Physical Properties Catalogue (MP3C) to include the available spectral classes of asteroids. Then, for each asteroid we selected the best diameter, rotational period, and spectral class. Building upon a semi-supervised machine-learning method, we quantify the valley between fast and slow rotators for S- and C-complex asteroids, which are known to be linked to different types of meteorites, namely ordinary and carbonaceous chondrites respectively. The method iteratively fits a linear boundary between the two populations in a rotational period-diameter space to maximise separation between the two populations.}
{We find a clear compositional dependence of the valley: for C-complex asteroids the transition occurs at longer periods than for S-complex, with
$P_{} = 14.4\,D_{\rm km}^{0.739}$ (C-complex) and $P_{} = 11.6\,D_{\rm km}^{0.718}$ (S-complex), where period and diameter are given in hours and kilometres respectively. This corresponds to $\mu Q \simeq 2$ and $13~\rm GPa$, respectively, where $\mu$ is the rigidity, measuring how strongly a body resists shear deformation under applied stress, and $Q$ is the quality factor, measuring how efficiently a body dissipates mechanical energy when cyclically deformed.}
{The dependence of the valley on spectral classes likely reflects compositional and structural differences: C-complex, being more porous and weaker, dissipate angular momentum more efficiently than stronger, more coherent S-complex. This represents quantitative evidence of class-dependent rotational valleys within asteroid populations.}


   \maketitle

\nolinenumbers

\section{Introduction}

Asteroids are remnants of the original planetesimals of the solar system, and they preserve important information about our origins. Studies of asteroid populations have become increasingly detailed as new datasets emerge, revealing complex spin-state behaviours \citep{durech2023}. Early work suggested that asteroid spin rates followed a Maxwellian distribution, consistent with a collisionally evolved population \citep{harris1979}. However, later observations on smaller asteroids (i.e. $<$ a few tens of kilometres) demonstrated that the distribution deviates significantly from this model, with two distinct groups: fast and slow rotators \citep{BINZEL1984294, pravec_2000_fast, pravec_2008_spin, durech2023}. Data from the ESA Gaia mission revealed a distinct “valley” in the period–diameter ($P$–$D$) diagram, separating slow and fast rotators \citep[see][and references therein]{durech2023}. A similar feature is also evident in Transiting Exoplanet Survey Satellite (TESS) observations \citep{Vavilov2025}. Although previous studies \citep{durech2023, zhou2025} referred to this feature as a “gap”, the term “valley” more accurately describes it, as it marks the boundary between two densely populated rotational regimes \citep{zhou2025}.

Despite decades of study, the rotational behaviour of asteroids reveals mysteries that are not understood by current models. In particular, why do some asteroids spin unusually slowly \citep{erasmus_2021_discovery}, or result in chaotic spin states, as we see in tumblers? What mechanisms, which are missing from our current models, maintain these two populations, and what determines the position of the valley that splits these two populations? This valley is not only observationally interesting, but could prove essential to understanding how their internal structure, density, and collisional history can shape the evolution of their rotational properties. Understanding these mysteries requires the separation of heavily entwined effects of torques, dissipations, and excitations which each act differently depending on the asteroids composition. Earlier work has suggested that these processes may operate with different efficiencies in different taxonomic classes, with C- and S-type asteroids exhibiting distinct transition diameters toward collisionally relaxed spin-rate distributions, interpreted in terms of composition-dependent YORP evolution \citep{carbognani_2011}.

The excess of fast rotators is well explained by the YORP (Yarkovsky-O’Keefe-Radzievskii-Paddack) effect, which is a torque due to the momentum carried away by thermal and reflected photons. The YORP effect can spin small bodies up to near their critical limits and maintain the spin rate by shedding mass into space \citep{rubincam_2000_radiative,walsh_2008_rotational}. Differences in bulk density and internal cohesion between taxonomic classes are therefore expected to influence these critical limits; in particular, C- and S-type asteroids have been reported to exhibit distinct spin-barrier values \citep{carbognani_2017}. Slow rotators, on the other hand, may not be straightforward to explain within simple, averaged YORP-based models \citep{pravec2018}. More detailed studies have shown that when YORP drives asteroids to states with very little angular momentum, they are easily influenced by small stochastic torques, which can excite non-principal-axis rotation, potentially trapping bodies in long-lived tumbling or cyclic spin states \citep{dvokrouhlick_2007_generalized, scical_2010_averaged, breiter_2011_yarkovskyokeeferadzievskiipaddack, breiter_2015_tumbling}. In this framework, the efficiency of internal dissipation becomes the fundamental controlling factor, where it can determine whether asteroids either quickly return to principal-axis rotation, or remain in chaotic states consistent with the interpretation of \cite{zhou2025}.

\cite{zhou2025} introduced a semi-supervised machine-learning approach that quantifies the valley between fast and slow rotators using a linear support vector machine (hereafter SVM). The SVM iteratively maximises the distance between the two populations in period-diameter space through stochastic descent, providing an objective method to quantify the boundary between these populations. These authors identified an excess of slow rotators in data released in Gaia Data Release 3, that cannot be explained by previous models of the YORP effect and collisional evolution. In particular, they reported a clear valley between fast and slow rotators in the observed distribution, and found that the majority of tumbling asteroids occupy the slow-rotator regime. To explain this observation, they proposed a rotational evolution model in which tumbling states arise from two processes: spin-down driven by the YORP effect, and excitation by collisions. In this framework, slow rotators are more prone to tumbling because their longer damping timescales and weakened YORP torques lead to slower rotational evolution. 

\citet{zhou2025} explained the excess of slow rotators as the accumulation of tumbling asteroids. The distribution of tumblers can be constrained by a transitional boundary where the internal damping energy equals the collisional excitation energy. Below this boundary, collisional excitation dominates, generating slowly rotating tumblers; above it, damping dominates, producing fast regular spinners \citep{pravec_2000_fast}. This transition corresponds to the  valley observed in the $P-D$ diagram between fast and slow rotators, and its quantitative description enables identification of the low-density region separating the two populations.

However, this study was performed on the total population of asteroids, regardless of their composition, which could potentially play a role in the transition from spinners and tumblers, and thus the position of the valley. Spectral classes offer a natural framework for investigating the physical basis of this valley, as they can be used to represent asteroid compositions. Then one can investigate how the valley might be affected by density and internal strength \citep{pravec_2008_spin}. The systematic shifts of the valley between classes could reveal how mechanical strength, porosity, and cohesion govern the evolution of the spin.

Nevertheless, large-scale taxonomic analyses have historically been challenging because relevant datasets are scattered across the literature. The Minor Planet Physical Properties Catalogue (MP3C)\footnote{\url{mp3c.oca.eu}} \citep{delbo2022} hosted at the Observatoire de la Côte d’Azur, is a relational database that consolidates data from several hundred literature sources. MP3C not only provides a unified repository for orbital and physical properties but also includes best values for several of these. In the MP3C database, the physical properties tables may contain multiple measurements for the same asteroid, as these values originate from different methods or publications. However, in the best value tables, a single representative value for each parameter and each asteroid is computed.
Until recently, the MP3C lacked information on the spectral classes of asteroids. Here, we present an extended literature search specifically targeting this parameter, compiling data from more than 250 spectral-classification sources. Once these data were available, we developed and applied a method to determine the best value of the spectral class for each asteroid.

In this work, our aim is to extend the work of \cite{zhou2025} by considering different spectral classes. We will be focusing on S and C taxonomic complexes, which are broadly linked to ordinary chondrites and the more primitive carbonaceous meteorites, respectively, that formed at different heliocentric distances. 

In Sec.~\ref{sec:data}, we discuss the assembly of this new dataset, to include spectral classification of asteroids into MP3C. In Sec.~\ref{sec:method} we discuss our target populations, and our methods used to study the dependence of the diameter on the rotational period, for the main two asteroid compositions. In Sec.~\ref{sec:resultsAndDiscussion} and~\ref{sec:conclusion}, we present our results and our discussion.

\section{Data}
\label{sec:data}

As of November 2025, MP3C hosts over 6.6 million measured properties (e.g. orbital elements, diameters, masses) for 1.46 million asteroids. Our objective is to introduce spectral classes and derive a single representative (`best') value for each physical property of the asteroids. 

For the quantitative parameters, namely diameter ($D$) and geometric visible albedo ($p_\text{V})$, we retrieved the values for the total asteroid population. In the case that there existed multiple values per asteroid, we used the best value that is defined as the uncertainty weighted average of the existing values. Similarly, for the rotational period and its quality, we use the best values given from MP3C, which are directly obtained from the Light Curve Data Base (LCDB) \citep{Warner2009}. Our dataset also includes rotational period data from \cite{durech2023} and \cite{athanasopoulos_2024_spin}, which are given a quality value of 2.

For asteroid spectral class data, which is a qualitative property in nature, we implemented a scoring-based approach to assign a best class to each asteroid. This method combines information from multiple measurements, accounting for the data collection method, wavelength coverage, and the taxonomic scheme used.
Each measurement is categorised according to the technique used for the acquisition of data: Spectroscopy (SPEC), spectrophotometry (PHOT), or a combination of both (SPEC + PHOT). In addition, the wavelength range covered was also used: visible (VIS), near-infrared (NIR), thermal-infrared (TIR) or a combination. A weighting scheme was used to assign values to different methods and wavelength ranges. This scheme assigns higher values to methods considered more informative, e.g., spectroscopic measurements generally receive higher weights than photometric ones, and measurements spanning broader wavelength ranges are favoured (Table~\ref{tab:measurement_weights}). Measurements are also weighted according to the classification system used (e.g., Tholen, Bus and Bus-DeMeo), giving more influence to modern, widely adopted schemes that are expanded into more wavelengths, while still considering older systems. We identified 17 taxonomic schemes for which we give weight=1, except the Bus-DeMeo widely-used taxonomy where we give weight=3 (Appendix~\ref{sec:appA}). 

For each asteroid, all available spectral class entries are aggregated. Each entry contributes a score computed as the product of its method/wavelength weight, taxonomic weight, and the reported probability of the class (if available in the original publication). The cumulative score is calculated separately for each class observed for that asteroid. The spectral class with the highest total score is assigned as the asteroid’s best class. This approach ensures that the resulting classification prioritizes high-quality, modern measurements while still incorporating historical data when newer observations are lacking. The scoring-based algorithm described here is the same procedure used operationally within MP3C to return the publicly available best spectral class when queried by users, ensuring full consistency between the catalogue and the present analysis.

\begin{table}[h!]
    \centering
    \caption{Weights for different combinations of wavelengths and methods of data collection.}
    \begin{tabular}{l c}
    \hline
    \hline
    Measurement & Weight\\
    \hline
    NIR and PHOT & 1 \\
    VIS and PHOT & 2 \\
    VISNIR and PHOT & 3 \\ 
    VISNIRTIR and PHOT & 4 \\
    VIS and SPEC & 5 \\
    NIR and SPEC & 6 \\
    VISNIR and SPEC & 7 \\
    VISNIRTIR and SPEC & 7 \\
    VIS and SPEC+PHOT & 8 \\ 
    NIR and SPEC+PHOT & 9 \\
    VISNIR and SPEC+PHOT & 10 \\
    VISNIRTIR and SPEC+PHOT & 10 \\
    \hline
    \end{tabular}
    \label{tab:measurement_weights}
\end{table}

Scores are computed per class per asteroid, sorted, and the top-scoring class is updated in the database for each object. This framework provides a systematic, reproducible method for synthesizing heterogeneous spectral measurements into a single representative classification. In Table~\ref{tab:spec_class} we show a representative example of this process using asteroid (161) Athor which belongs to a well characterised spectroscopic class \citep{avdellidou2022}. Following the method set out, we determined that the most likely (and therefore best) spectral class is Xc, with a score of 52. When this method is applied to our whole asteroid population, it helps us significantly reduce the complexity of our dataset, enhancing the usability of the dataset. Throughout this work, this best spectral class is the one adopted for population selection, filtering, and all subsequent analyses.

\begin{table}[ht]
    \centering
    \caption{Table showing the spectral classification of Athor (161), as taken from mp3c.oca.eu.}
    \begin{tabular}{llll}
        \hline
        \hline
        Spectral Class & Taxonomy  & Wavelength & Method \\ 
        \hline
        A               & Mahlke    & VIS       & SPEC \\ 
        M               & Tholen    & VIS       & PHOT \\ 
        M               & Mahlke    & VISNIR    & SPEC \\ 
        Xc              & Bus-DeMeo & VISNIR    & SPEC \\ 
        Xc              & Bus       & VIS       & SPEC \\ 
        Xc              & Bus       & VIS       & SPEC \\ 
        Xc              & Bus-DeMeo & VISNIR    & SPEC \\ 
        \hline
    \end{tabular}
    \label{tab:spec_class}
\end{table}

\begin{figure}
    \centering
    \includegraphics[width=\linewidth]{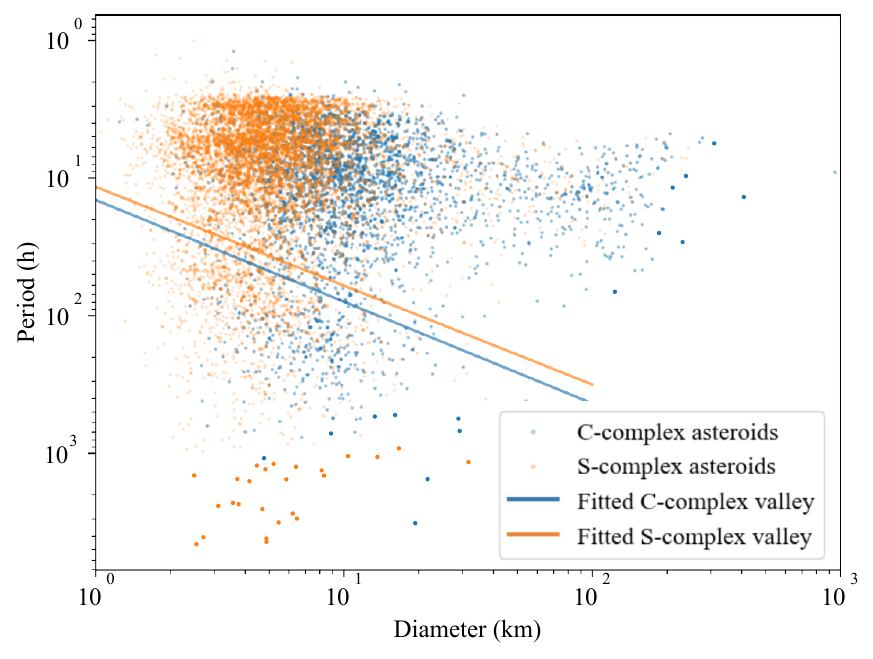}
    \caption{Period-diameter distribution and the fitted valley functions for S- and C- complex asteroid. }
    \label{fig:data}
\end{figure}

\section{Methods}
\label{sec:method}

\subsection{Population selection}
\label{sec:selection}
From the population of asteroids that has a recorded diameter, geometric visible albedo, and rotational period, we selected those that belong to the two main spectroscopic groups of the Bus-DeMeo taxonomy, namely, the S-complex (S, Sa, Sq, Sr, and Sv classes), and the C-complex (B, C, Cb, Cg, Cgh, and Ch classes). X-complex asteroids are not analysed in this work because this spectral group encompasses multiple distinct compositions, and the limited number of objects with well-determined subclasses prevents a meaningful compositional separation.

Each taxonomic group was further restricted using geometric visible albedo values, adopting a commonly used threshold of $p_V = 0.12$ \citep{delbo_2017_identification} to separate low-albedo C-complex objects ($p_V \leq 0.12$) from higher-albedo S-complex objects ($p_v>0.12$) \citep{mainzer_2011_neowise, masiero2011, demeo_2013_the}. Objects whose best spectral classification was inconsistent with this albedo criteria (e.g. C-complex with $p_V$ greater than 0.12) were excluded to minimise contamination from ambiguous or uncertain classifications.

We also filtered our population to only include asteroids where their period quality value was at least 2- (2-, 2, 2+, 3-, 3) from LCDB \citep{Warner_2023}.
This selection process resulted in 6049 S-complex and 2947 C-complex objects, as presented in Fig.~\ref{fig:data}.

Further filtering regarding the YORP timescale of objects was also considered. Adopting a nominal YORP timescale of $\sim$1 Myr for a 1-km body, and the standard scaling $\tau_\mathrm{YORP} \propto D^2$ \citep{jewitt_2025_nongravitational}, a Solar System age of 4.5 Gyr corresponds to $D$ $\simeq$ 65 km. Using the collisional lifetimes in the main belt from \citeauthor{marchi_2006_a} (\citeyear{marchi_2006_a}; based on \citeauthor{bottke2005} \citeyear{bottke2005}), YORP evolution becomes comparable to or slower than catastrophic disruption timescales for $D \gtrsim$ 50 km. Removing such large objects from our sample has negligible impact on the inferred valley slope, owing to their small number and limited statistical weight. We nevertheless retain $D \geq 50km$ in our analysis for two reasons. First, they offer a useful comparison between the spin distributions of small, YORP-evolved asteroids and large bodies whose rotational states may remain essentially primordial. Second, the effective gigayear-scale YORP timescale remains uncertain, particularly due to stochastic reorientation by non-catastrophic collisions, so that the classical static estimate likely underestimates the long-term evolution time.

\subsection{Quantifying the boundary of slow and fast rotators}
\label{sec:ML}

To identify the valley between slow and fast rotators in both spectral complexes, we adopt the method of \citet{zhou2025}. This approach employs a semi-supervised machine-learning algorithm that classifies asteroids into two groups and determines the boundary line separating them, which represents the centre of the valley. In their procedure, pseudo-labelling was first performed by manually defining an unclassified “grey zone” as a reference region. The grey zone is centred on the presumed location of the valley, with its midpoint given by the reference line and its width corresponding to the distance along the y-axis from that line. Data points above this region were assigned to populations above the valley (spinners), and those below the valley (tumblers).

The reference line (i.e., the valley) and the associated grey zone were then iteratively updated by fitting the data, using SVM, to maximize the separation between the two subpopulations. In logarithmic coordinates, the valley is expressed
\begin{equation}
\label{eq:valley_function}
    \log P_{\rm h} = k \log D_{\rm km} + b
\end{equation}
where the subscripts indicate the units. Through repeated updating of the grey zone and refitting, the model converges, and the reference line is identified as the functional form of the valley. 

To account for uncertainties in asteroid diameters, we generated 1,000 synthetic datasets for both the S- and C-complex populations, assigning each asteroid a diameter drawn from a Gaussian distribution centred on its measured value, with a standard deviation equal to its quoted uncertainty. 
In addition, to test the robustness of our results against the choice of the spectral score, we determined the values of k and b for different spectral score thresholds, between 1 (the nominal value) and 5. Increasing the spectral score restricts our dataset to higher quality data, while decreasing the number of data points (Appendix~\ref{sec:appB}).

\section{Results and discussion}
\label{sec:resultsAndDiscussion}

\begin{figure}
    \centering
    \includegraphics[width=\linewidth]{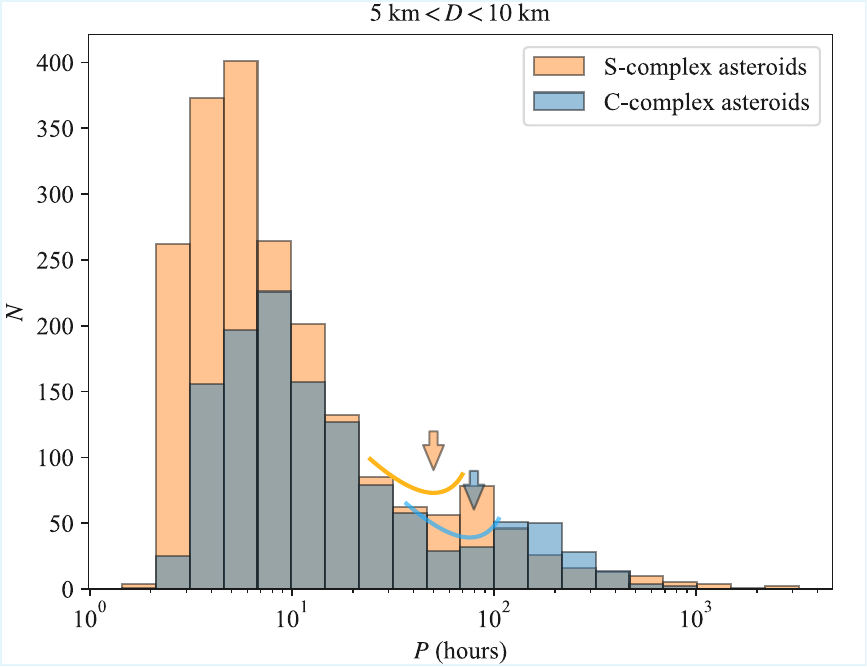}
    \caption{Spin-period distribution of S-complex and C-complex asteroids in 5-10~km diameter range. The local minimum in each distribution corresponds to the valley of the $P$-$D$ diagram. The minimum for C-complex asteroids (blue) occurs at longer periods than for S-complex asteroids (orange).}
    \label{fig:period_histogram}
\end{figure}

\begin{figure}
    \centering
    \includegraphics[width=\columnwidth]{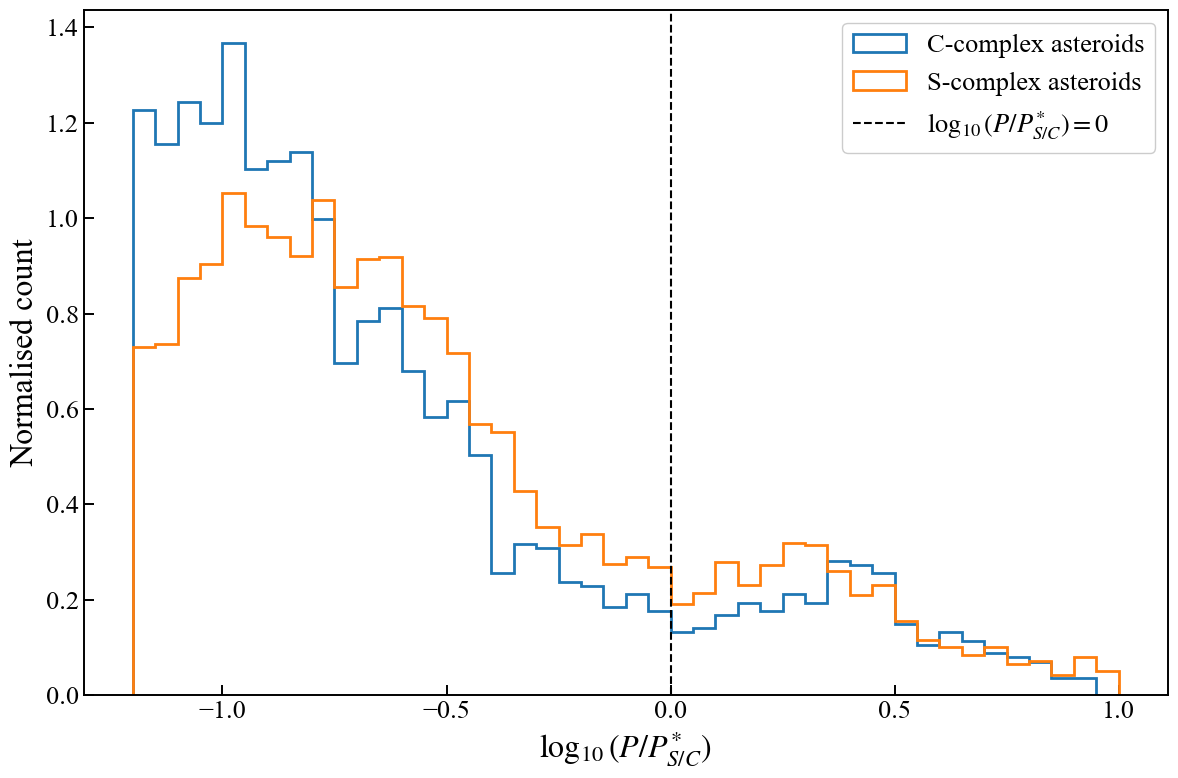}
    \caption{Normalised distributions of the dimensionless offset from the fitted valley loci, $\log_{10}(P/P^{*}_{\scriptscriptstyle S})$ and $\log_{10}(P/P^{*}_{\scriptscriptstyle C})$, shown separately for the S-complex (orange) and the C-complex (blue). The vertical dashed line marks $\log_{10}(P/P^{*}_{\scriptscriptstyle S}) = 0$ and $\log_{10}(P/P^{*}_{\scriptscriptstyle C})=0$, corresponding to the valleys' centre.}
    \label{fig:ppstar_histogram}
\end{figure}
 
\begin{figure}
    \centering
    \includegraphics[width=\linewidth]{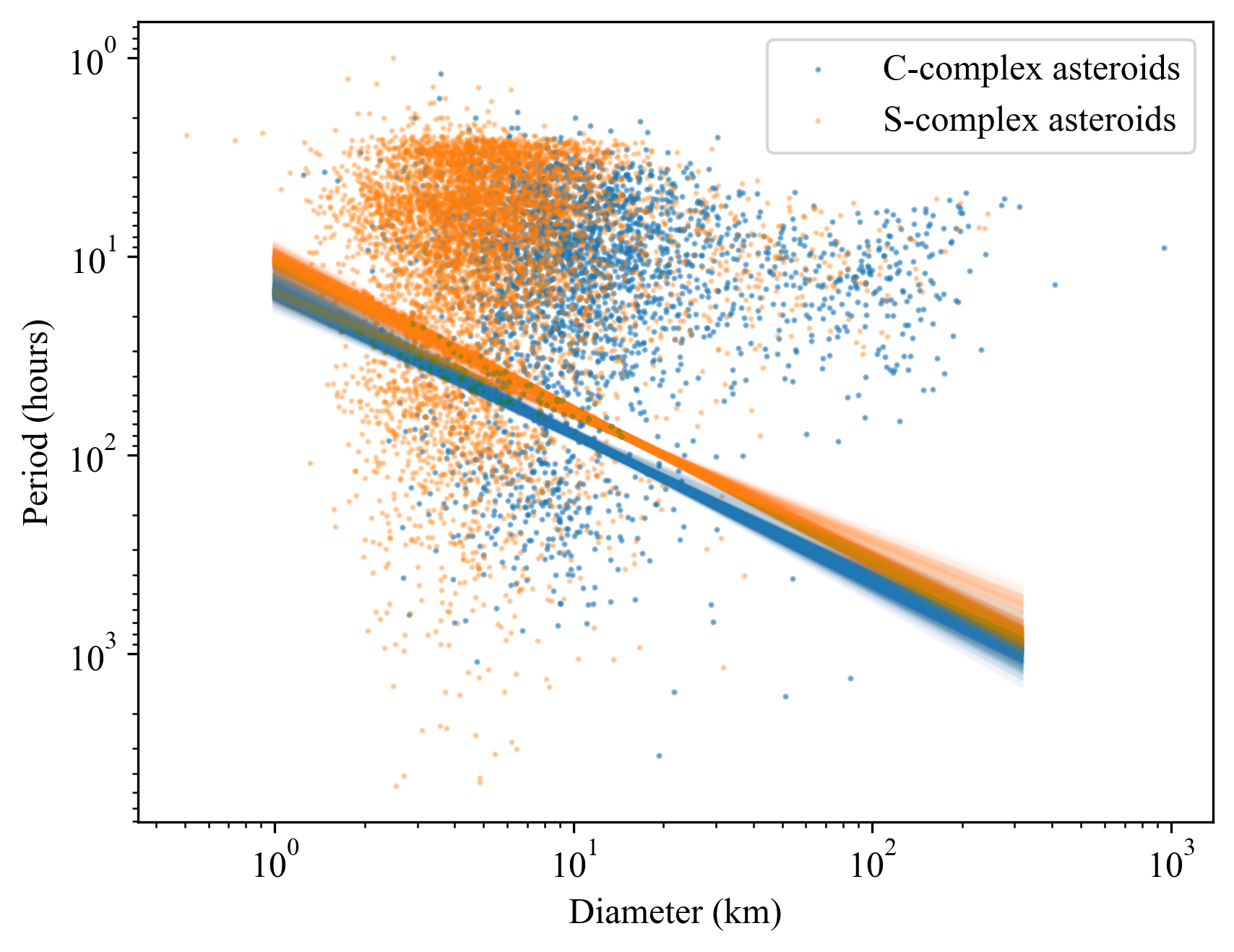}
    \caption{Period–diameter ($P-D$) distribution for S-complex and C-complex asteroids. The lines are fitted functions for the S-complex valley and C-complex valley in the synthetic datasets generated according to the diameter uncertainty.}
    \label{fig:P_D_synthetic_diagram}
\end{figure}

\begin{figure}
    \centering
    \includegraphics[width=\linewidth]{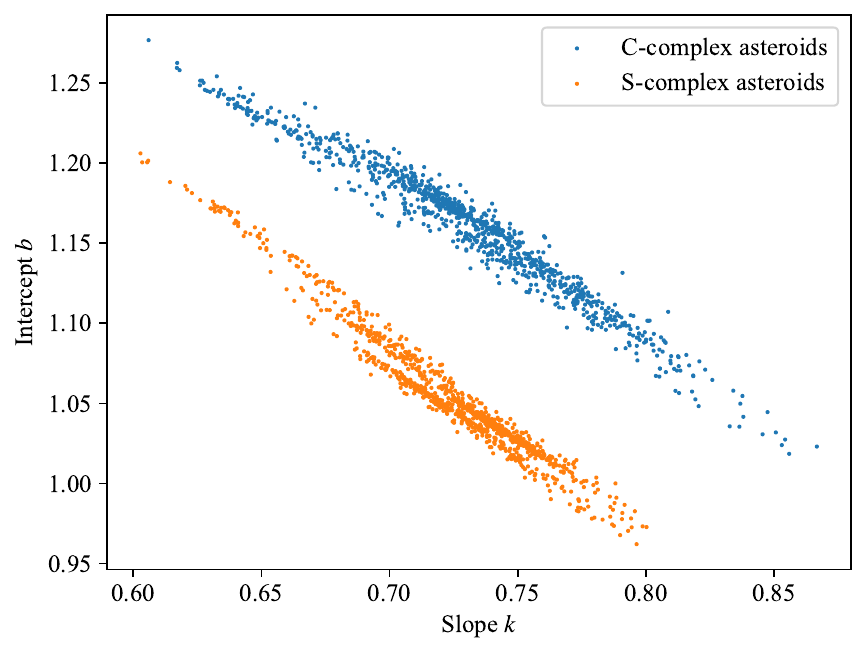}
    \caption{Distributions of the valley parameter $k$ and $b$ (Eq.~\ref{eq:valley_function}) for S-complex and C-complex asteroids, derived from fits to synthetic datasets that account for diameter uncertainties. Larger values of $b$ correspond to a lower vertical position of the valley in the $P$–$D$ diagram. }
    \label{fig:k_b_distribution}
\end{figure}

\subsection{Distinct valley locations for S-complex and C-complex}

Using data selected as described in Sec.~\ref{sec:data} and \ref{sec:selection}, first, we plot the spin period histogram for S- and C-complex asteroids in a diameter range between 5 and 10~km (Fig.~\ref{fig:period_histogram}). This size range is chosen because it contains sufficient data and falls in a regime where the YORP effect is effective \citep{rubincam_2000_radiative}.
Outside this range the valley is expected to be absent or much less pronounced because smaller bodies are less well sampled, whereas larger ones experience weaker YORP torques.

The histogram of Fig.~\ref{fig:period_histogram}, which corresponds to the integrated valley region in the $P$–$D$ diagram, show clear local minima, indicated by the vertical arrows. These minima mostly occur at shorter spin periods for S-complex asteroids than for C-complex ones.
The offset between the local minima for the two distinct spectroscopic groups of asteroids indicates that the valley of the S-complex population lies higher, i.e., towards shorter periods in the $P$–$D$ plane, than the valley of the C-complex (Fig.~\ref{fig:data}).

To quantify this difference, and its dependence with asteroids' diameters, we applied the fitting method described in Sec.~\ref{sec:ML}. The results are
\begin{equation}
\label{eq:S_valley}
    P^{*}_{\scriptscriptstyle S} = 
    11.6 {\rm ~h} \left( {D \over {1 ~\rm km}}\right)^{0.718}
\end{equation}
for S-complex asteroids and
\begin{equation}
\label{eq:C_valley}
    P^{*}_{\scriptscriptstyle C} =
    14.4 {\rm ~h} \left( {D \over {1 ~\rm km}}\right)^{0.739}
\end{equation}
for C-complex asteroids, as shown in Fig.~\ref{fig:data}, where the $P^{*}_{\scriptscriptstyle S/C}$ symbols indicate the critical $P$-value as a function of $D$ for each asteroid compositional complex, namely, $P^{*}_{\scriptscriptstyle S/C} \equiv P^{*}_{\scriptscriptstyle S/C}(D)$. Because Eq.~\ref{eq:S_valley} and ~\ref{eq:C_valley}, represent distinct functions, this is the first clear, quantitative evidence that the location of the valley depends on spectral class. 

The functions $P^{*}_{\scriptscriptstyle S}(D)$ and $P^{*}_{\scriptscriptstyle C}(D)$ describe both the position and the slope of the valley as functions of the asteroids' sizes for each spectral complex.

For each complex separately, by dividing each asteroid’s  rotation period $P$ by the corresponding valley values, $P^{*}_{\scriptscriptstyle{S/C}}$, we obtain the dimensionless ratio $P/P^{*}_{\scriptscriptstyle{S/C}}$, which measures the offset of an asteroid from the valley. 
Objects with $P/P^{*}_{\scriptscriptstyle{S/C}} < 1$ rotate faster than those at the centre of the valley, while those with $P/P^{*}_{\scriptscriptstyle{S/C}} > 1$ rotate more slowly. In this representation, the valley appears as a deficit of objects near $P/P^{*}_{\scriptscriptstyle{S/C}} \simeq 1$.
Fig.~\ref{fig:ppstar_histogram} shows the distribution of the normalized rotational periods, $P/P^{*}_{\scriptscriptstyle S/C}$ of the asteroids of the S-complex and the C-complex, for all values of their diameters. Both histograms clearly show the valley between fast and slow rotators, further corroborating the presence of the valley.

We conclude that in $P-D$ space (Figure~\ref{fig:data})  the S-complex valley is higher than C-complex valley. While the fit suggests that two complexes may differ in both offset and slope, when we apply the uncertainty estimation method described in sec. \ref{sec:ML}, we find the slope differences is not statistically significant, given the current sample size and diameter uncertainties (Fig.~\ref{fig:k_b_distribution}). So we cannot conclude whether the contrast between S- and C-complex asteroids is also shown in a slope difference. On the other hand, what is robust across the tests performed is that the S-complex valley lies above the C-complex valley. The resulting fitted valley functions are shown in Fig.~\ref{fig:P_D_synthetic_diagram}, clearly demonstrating that the S-complex valleys generally lie above those of the C-complex. The figure visually shows that the uncertainty is the smallest for $D$ between $\sim$10 and 20 km, due to the smaller relative diameter uncertainties at these sizes.
Fig. \ref{fig:k_b_distribution} results from the application of our method to estimate the uncertainties in the valley parameters, $k$ and $b$ (sec. \ref{sec:method}) and shows the distinct distributions of these parameters  (Eq.~\ref{eq:valley_function}) for S- and C-complex asteroids.  Larger values of $b$ correspond to lower locations in the $P$–$D$ diagram. The S-complex parameters lie to the bottom of those of the C-complex, further demonstrating that the S-complex valley is systematically higher. Our methods allows us to conclude that  our results are robust against uncertainties in the observational data. In addition, we tested the robustness of our results as function of the adopted threshold for the spectral score, which affects the selection of the asteroids used in the study. Fig. \ref{fig:spectral_robustness} shows the values of valley parameters $k$ and $b$ when we apply our method with different spectral score thresholds. Because the results lie within the distribution of the $k$ and $b$ parameters when we take into account the uncertainties in the data, we deduce that our results are robust against selection applied by the spectral score threshold.

In the work of \citet{zhou2025}, the valley is found to have a strong correlation with distribution of tumblers in $P-D$ diagram, and thus explained as the separatrix between spinners and tumblers, which is relevant to the location where collisional excitation and frictional damping is balanced. The timescale of collisional excitation for an asteroid with the bulk density of $\rho$ can be estimated as
\begin{equation}
\label{eq:tau_col}
    \tau_{\rm col} \sim 113~{\rm Myrs} \, \left( {D \over 1~{\rm km}} \right)^{(4\alpha - 10) / 3} \left( {P \over 8{\rm ~h}} \right)^{(1-\alpha)/3} \left( \frac{\rho}{2~{\rm g~cm^{-3}}} \right)^{(\alpha - 1)/3}  
\end{equation}
using the standard estimation of collisional probability for asteroids \citep[see details in][]{Farinella1998,zhou2025}. The dependence of Eq.~\ref{eq:tau_col} on bulk density was not explicitly shown in \citet{zhou2025}, as that work assumed equal densities for the impactor and the target asteroid. In the present study, we adopt a mean bulk density of $2~\mathrm{g~cm^{-3}}$ for the impactors, while the bulk densities of S- and C-complex asteroids are taken to be $3$ and $1.5~\mathrm{g~cm^{-3}}$, respectively. Here $\alpha$ is the power index of the differential size-frequency distribution of asteroids that sets the number of impactors. 

On the other hand, the timescale of frictional damping depends on the material parameter $Q/k_2$. Tumbling causes the asteroid to undergo time-varying internal deformation as the orientation of the rotation axis continuously changes relative to its principal axes \citep[e.g.,][]{Burns1973,Harris1994,Pravec2014}. The efficiency of this deformation is described by the degree-2 Love number $k_2$, which quantifies the amplitude of the induced gravitational potential relative to the external forcing. A small $k_2$ indicates that the body resists deformation and stores only a small amount of elastic strain energy. For a homogeneous elastic body, $k_2$ is determined by the competition between self-gravity and internal rigidity. In the limit where rigidity dominates over self-gravity ($\mu \gg \rho gR$), $k_2$ can be expressed as
\begin{equation}
\label{eq:k2_1}
    k_2 \sim \frac{\pi G \rho^2 D^2}{19 \mu}.
\end{equation}
where $\mu$ is the shear modulus (rigidity) characterizing the elastic resistance to shear deformation. This relation shows that small bodies (e.g., rubble-pile asteroids) possess extremely small $k_2$ due to both their weak self-gravity and low effective rigidity. The stored strain energy is partially dissipated as heat through internal friction, quantified by the tidal quality factor $Q$. The parameter $Q$ measures the ratio of the elastic energy stored to that lost per cycle of deformation, so a lower $Q$ denotes stronger dissipative processes. Energy dissipation is quantified by the tidal quality factor $Q$, defined as the ratio of the elastic energy stored to the energy lost per cycle of deformation. A high $Q$ corresponds to weak dissipation (nearly elastic response), while a low $Q$ indicates strong internal friction and rapid energy loss.

Therefore, a high $Q/k_2$ or $\mu Q$ represents longer damping timescale:
\begin{equation}
\label{eq:tau_damp}
    \tau_{\rm damp} \simeq A{\mu Q\over 2 \pi^3\rho } P^3 D^{-2} ,
\end{equation}
where $A \sim 18$ is a dimensionless shape factor \citep{Breiter2012, Pravec2014}. Large $\mu Q$ corresponds to weak energy loss and slow damping, whereas small $\mu Q$ produces rapid relaxation toward principal-axis rotation.
\citet{zhou2025} provide the normalized form:
\begin{equation}
\label{eq:tau_damp_norm}
    \tau_{\rm damp} \simeq  0.1~{\rm Myrs} ~{\left(\mu Q \over 10^{9}{\rm Pa} \right)}~{\left( \rho \over 2 {\rm ~g~cm^{-3}} \right)^{-1}} \left( {D \over 1~{\rm km}} \right)^{-2} \left( {P \over 8{~\rm h}} \right)^3
\end{equation}

When equating the excitation and damping timescales (Eqs.~\ref{eq:tau_col} and \ref{eq:tau_damp_norm}), we find the condition where these two processes balance:
\begin{equation}
\begin{aligned}
    P &= 8 {\rm ~h} \cdot~283^{3/(\alpha+8)} \left( {\mu Q \over 4 \times 10^9{\rm ~Pa}} \right)^{-{3 / (8+\alpha})} \left( { \rho \over 2~{\rm g~cm^{-3} }} \right)^{{(\alpha+2) /( \alpha + 8)}} \\
    & \cdot \left( {D \over 1{\rm ~km}} \right)^{(4\alpha -4) / (8+\alpha)}.
\end{aligned}
\end{equation}
This sets the boundary of the distribution of tumblers. The centre of the valley that is detected using the machine learning method is a bit higher than it:
\begin{equation}
\begin{aligned}
    P^{*} &= 8 {\rm ~h} ~\cdot 7^{3/(\alpha+8)}~\left( {\mu Q \over 4 \times 10^9{\rm ~Pa}} \right)^{-{3 / (8+\alpha)}} \left( { \rho \over 2~{\rm g~cm^{-3} } } \right)^{{(\alpha + 2) / (\alpha+8)}} \\
    & \cdot \left( {D \over 1{\rm ~km}} \right)^{(4\alpha -4) /(8+\alpha)}.
\end{aligned}
\label{eq:valley}
\end{equation}
\citet{zhou2025} found that, when assuming $\alpha = 3.2$, a global fit to all asteroids requires $\mu Q$ of $ 4 \times10^9$ Pa, which is several orders of magnitude below monolithic values ($10^{11}$–$10^{13}$ Pa).
Comparing Eqs.~\ref{eq:S_valley}, \ref{eq:C_valley} and \ref{eq:valley}, we find $\alpha \sim 3$ for both S- and C-complex asteroids. Adopting $\rho = 1.5~\mathrm{g\,cm^{-3}}$ for C-complex asteroids and $\rho = 3~\mathrm{g\,cm^{-3}}$ for S-complex asteroids \citep{Fienga2020}, we obtain $\mu Q \sim 1.3 \times 10^{10}~\mathrm{Pa}$ for S-complex and $\mu Q \sim 2 \times 10^{9}~\mathrm{Pa}$ for C-complex asteroids. This is consistent with the $\mu Q \sim 4 \times 10^9~$Pa for the whole asteroid population found by \citet{zhou2025}.

Following \citet{Breiter2012}, we adopt a phenomenological $Q$-model in which the quality factor is treated as frequency-independent. While more realistic rheological models may predict a frequency dependence of $Q$, such effects are poorly constrained for asteroid materials and are commonly neglected in analytical treatments of wobble damping. Importantly, our analysis focuses on relative dissipation efficiencies between asteroid classes; thus, any similar frequency dependence of $Q$ across compositions would not affect our qualitative conclusions.

\subsection{Implications on internal structure and material properties}

The offset between the S- and C-complex valleys gives evidence that the balance between collisional excitation and frictional damping differs for S- and C- complex asteroids through $\mu Q$. The slower transitional periods (lower valleys) of C-complex asteroids indicate stronger damping and therefore smaller $\mu Q$, consistent with higher porosity, weaker cohesion, and greater energy loss per cycle. In contrast, the higher valleys of S-complex asteroids imply more elastic, less dissipative interiors. These denser, silicate-dominated aggregates require faster spins to damp their tumbling motion (e.g. $\tau_{\rm damp} \propto P^3$ in Eq.~\ref{eq:tau_damp}) before their rotation is re-excited by collisions, whereas carbonaceous C-complex bodies more readily deform and dissipate rotational energy, allowing them to maintain principal-axis rotation at lower spin rates. Laboratory tensile-strength measurements of carbonaceous meteorites support this interpretation, showing strengths an order of magnitude lower than those of ordinary chondritic, silicate-rich analogues \citep{pohl_2020_strengths}.

The parameter $\mu Q$ also depends on near-surface structure. In the model of \citet{Nimmo2019}, a thicker regolith layer leads to more energy dissipation and hence a smaller $\mu Q$. Regolith is shaped by surface activity such as space weathering effects, thermal fatigue \citep{delbo2014} and internal processes such as landslides, seismic shaking, and creep \citep[e.g.][]{Zhang2018,Brisset2018,Cheng2021}. However, it is important to note that the “regolith’’ implied in the $Q/k_2$ framework of \citet{Nimmo2019} does not necessarily correspond to the near-surface regolith usually discussed in thermal models of small asteroids. In the $Q/k_2$ formulation, the dissipative layer is the portion of the body that contributes to tidal energy dissipation, and its characteristic thickness can reach tens of meters or more \citep{Nimmo2019, Pou2024}. By contrast, in thermal contexts the regolith thickness is typically assumed to be of order a few thermal skin depths (i.e. a few centimetres), and S-type asteroids are in fact expected to generate thicker near-surface regolith layers than C-complex asteroids \citep{Cambioni2021}.

Beyond rotation, $\mu Q$ governs the efficiency of tidal evolution in binary asteroids. Roughly 15\% of asteroids are binaries \citep{ppravec_2006_photometric} whose long-term evolution is driven by tidal torques inversely proportional to $\mu Q$ \citep[e.g.][]{Murray1999}. A lower $\mu Q$ for C-complex bodies implies stronger tidal dissipation and faster orbital evolution, potentially contributing to the observed deficit of C-complex binaries \citep{Minker2023, liberato_2024_binary}.

It is important to note that the physically meaningful quantity is the dimensionless ratio $Q/k_2$, whose modelling remains uncertain. Equation~\ref{eq:k2_1} is based on an elastic-sphere assumption, which is the not the case of rubble-pile asteroids. Therefore, the rigidity $\mu$ in Eq.~\ref{eq:k2_1} is not a true material modulus but an effective parameter. The mapping from $Q/k_2$ to $\mu Q$ and ultimately to internal structure is a degenerate problem at the current stage. We would like to briefly review other models of $Q/k_2$ from Eq.~\ref{eq:k2_1} to illustrate how $Q/k_2$ (or $\mu Q$) reflects the internal structures and material properties of rubble piles.

\citet{Goldreich2009} first derived $k_2$ of a rubble pile by considering void-mediated yielding within a granular aggregate. They found
\begin{equation}
\label{eq:k2_2}
    k_2 \sim \rho R \sqrt{G \epsilon_{\rm Y} \over \mu_{\rm r}}
\end{equation}
where $\epsilon_{\rm Y}$ is a characteristic yield strain ($\sim 10^{-2}$) and $\mu_{\rm r}$ is the effective rigidity for rubble piles. The damping timescale is thus
\begin{equation}
\label{eq:tau_damp_2}
    \tau_{\rm damp} = {Q  \over 19\pi^2 } \sqrt{{G\mu_{\rm r} \over  \epsilon_{\rm Y}}} {P^3 \over D},
\end{equation}
which yields a valley function (combining Eq.~\ref{eq:tau_col} and \ref{eq:tau_damp_2})
\begin{equation}
\label{eq:P_D_2}
    P \propto   \left( \sqrt{\mu_{\rm r}} Q \right)^{-3/(8+\alpha)}  D^{(4\alpha -7)/(8+\alpha)}
\end{equation}.
This model predicts a valley slope $k = (4\alpha - 7)/(8+\alpha) \simeq 0.51 $, smaller than the observed $0.60$–$0.85$ \citep[][and Fig.~\ref{fig:k_b_distribution}]{zhou2025}, which may imply that additional size-dependent dissipation must be present.

The quality factor may also have size dependence for rubble piles. \citet{Nimmo2019} developed a model in which energy loss only occurs within a surface regolith (a dissipative layer) of thickness \textit{h}. The resulting effective quality factor scales inversely with the square of $h$: thinner, looser surface layers dissipate more efficiently. The approximate scaling of the quality factor $Q$ is
\begin{equation}
    Q \propto \left(  {D\over h}\right)^2,
\end{equation}
which indicates, for a given regolith thickness $h$, $Q$ increases with the diameter $D$.  This trend arises because larger bodies experience smaller surface strain per unit stress. A rubble pile with a thick (tens-of-meters) regolith will therefore exhibit a larger $Q/k_2$ and hence a larger effective $\mu Q$ than a smaller object. The resulting $Q/k_2 \propto D$ is consistent with the trend in \citep{Jacobson2011} obtained by assuming binary asteroids are in balance of the tidal effect and the Binary YORP effect. \citet{Pou2024} exploited the ages of asteroid pairs to place quantitative bounds on $Q/k_2$ for both primaries and secondaries in 13 binary systems. They constrained $Q/k_2$ ranges from $10^4$ to $10^6$ for asteroid (3749) Balam with $D \sim 2~$km, with speculation of the regolith height $h$ ranging from $30~$m to 100~m. Translating to $\mu Q$ prescription using Eq.~\ref{eq:k2_1}
, we obtain $\mu Q \sim 10^7-10^9~$Pa, far lower than the value for monolithic bodies. However, these $Q/k_2$ values may still be uncertain because previous analyses neglected the recently proposed Binary Yarkovsky (BYarkovsky) effect \citep{Zhou2024a,Zhou2024b}, an eclipse-driven thermal recoil that can contract binary orbits. Incorporating this additional torque will be essential for future, self-consistent derivations of $Q/k_2$ and $\mu Q$.

In summary, the upward displacement of the S-complex valley relative to the C-complex valley in the $P$–$D$ diagram can be traced to a higher $\mu Q$, reflecting more rigid and less dissipative material behaviour. This compositional dependence connects the rotational dynamics of single asteroids with the tidal and thermal evolution of binary systems, providing a unified diagnostic of internal structure and surface properties across the asteroid population.

Ultimately, the present analysis provides observational evidence that asteroid spin-rate distributions are governed by a nuanced interplay between composition, internal structure, and evolutionary timescales. Accounting for these factors is essential for both modelling spin evolution and assessing the mechanical response of near-Earth objects in planetary defence contexts.

Looking ahead, these results also have implications for planetary defence applications. The same physical processes shaping the spin-rate distributions of main-belt asteroids operate in Near Earth Objects (NEOs) and Potentially Hazardous Objects (PHOs), where material properties and internal structure help advise impact-risk assessments. As forthcoming surveys such as the one from the LSST, and more in the future by NASA's NEOs Surveyor and possibly by ESA's NEOMIR space telescopes, dramatically will increase the sample of NEOs/PHOs with well-measured lightcurves, extending the present analysis to smaller and potentially hazardous bodies will enable tighter constraints on their bulk strength and spin-dependent failure modes. These parameters are central to evaluating mitigation strategies. Moreover, because the Yarkovsky effect is intrinsically linked to an asteroid’s rotation state, a refined understanding of spin evolution directly improves predictions of long-term orbital drift, thereby enhancing our ability to model, track, and forecast the trajectories of objects that may pose a threat to Earth.
\section{Conclusions}
\label{sec:conclusion}
By combining MP3C data with the semi-supervised method of \citet{zhou2025}, we have shown that asteroid rotation distributions differ between the two major spectral complexes. Both populations show similar patterns with a measurable deficit of objects in intermediate periods forming the valley in the Period-Diameter ($P-D$) diagram we described earlier. Crucially, the C-complex valley occurs at longer periods than that of the S-complex, reflecting intrinsic compositional and structural differences between the spectral classes. This demonstrates that the spectral information can play a pivotal role in interpreting the rotational evolution of asteroids, where some of these trends may not be as apparent in wider population studies.

These results show the potential of MP3C as a relational database being used as a tool for population-level asteroid studies, especially for the evolutionary processes of small-bodies. Further progress in this research will come from extending this analysis into individual subclasses within the complexes, or perhaps analysis of rarer spectral types (V-, T-, D-class asteroids), although both of these suggestions suffer from smaller sample sizes, which may require the adaption of current methods, alternative methods, or the collection of more data. As the dataset expands with new findings from the LSST and Gaia Data Release 4, this should provide the leverage needed to undergo these other test cases, and also test the robustness of these findings to better understand the underlying mechanisms that are responsible for the differences we see in rotational across our populations.

\begin{acknowledgements}
This work is based on data provided by the Minor Planet Physical Properties Catalogue (MP3C; mp3c.oca.eu). T. J. Dyer acknowledges financial support from French Space Agency CNES and the Université Côte d'Azur (UniCA). W.-H. Zhou acknowledges support from the Japan Society for the Promotion of Science (JSPS) Fellowship (P25021). M. Delbo also acknowledges support from CNES. The work of J. \v{D}urech was supported by the grant 23-04946S of the Czech Science Foundation.
\end{acknowledgements}

\bibliographystyle{aa}
\bibliography{references.bib}

\begin{appendix}
\section{References to taxonomic schemes}
\label{sec:appA}
Taxonomic schemes used in MP3C:

Barucci: \citep{mantoniettabarucci_1987_classification}

Binzel: \citep{binzel_1993_chips}

Birlan: \citep{mirelbirlan_1996_effects}

Bus: \citep{bus_2002_phase}

Bus-DeMeo: \citep{demeo_2009_an}

Carvano: \citep{carvano_2010_sdssbased}

Dahlgren-Lagerkvist: \citep{dahlgren_1999_a}

DeMeo-Carry: \citep{demeo_2013_the}

Fornasier: \citep{sfornasier_2011_spectroscopic}

Gradie-Tedesco: \citep{tedesco_1987_discovery}

Howell: \citep{howell_1994_classification}

Jewitt-Luu: \citep{jewitt_1990_ccd}

Mahlke: \citep{mahlke_2022_asteroid}

Popescu: \citep{mansour_2019_distribution}

Tholen: \citep{tholen_2017_asteroid}

\section{Spectral Score Robustness}
\label{sec:appB}
\FloatBarrier
The spectral classification adopted in this work relies on a scoring-based aggregation of different taxonomic measurements, which necessarily involves a trade-off between sample size and classification reliability. To verify that our main results are not driven by the particular choice of spectral-score threshold used to define the working sample, we explicitly test the robustness of the fitted valley parameters to progressively stricter spectral-quality cuts. This appendix illustrates how the fitted valley solutions for the S-complex and C-complex populations evolve as the minimum spectral score requirement is increased.
\begin{figure}[ht]
    \centering
    \includegraphics[width=\linewidth]{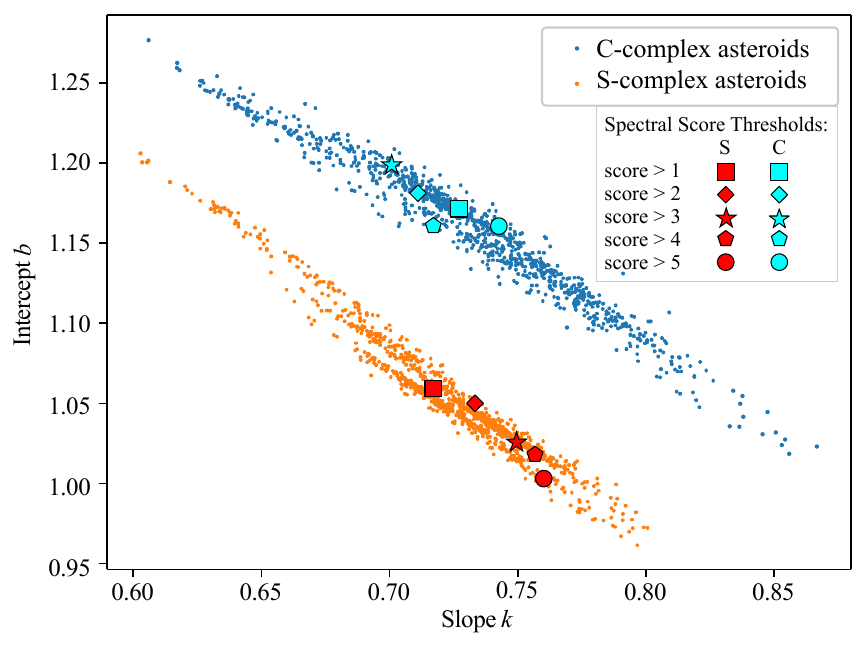}
    \caption{Same as Fig. \ref{fig:k_b_distribution}, showing the distributions of the valley parameters $k$ and $b$ (Eq.~\ref{eq:valley_function}) for S-complex and C-complex asteroids. Overplotted points indicate the best-fit valley solutions obtained from the dataset using progressively stricter spectral-classification score thresholds between 1 and 5 (as indicated by different symbols). Because the symbols lie within the distributions from the nominal datasets (see text), this illustrates the robustness of the fitted parameters versus changes in the spectral score threshold to the adopted spectral-quality cut.}
    \label{fig:spectral_robustness}
\end{figure}

\end{appendix}

\FloatBarrier 
\clearpage

\end{document}